\documentclass[%
journal=apl,
manuscript=letter,
doi=true,
layout=twocolumn
aip,
showpacs,
amsmath,amssymb,
reprint,%
]{revtex4-1}

\usepackage{graphicx}% Include figure files
\usepackage{dcolumn}% Align table columns on decimal point
\usepackage{bm}% bold math
\usepackage[usenames, dvipsnames]{color}

\def\Mo*{MoS$_2$}
\newcommand*\unit[1]{\,\mathrm{#1}} % for units in math mode

\begin{document}

\title{ \color{BlueViolet} Control of excitons in multi-layer van der Waals heterostructures}

\author{E. V. Calman}
\email{ecalman@gmail.com}
\author{C. J. Dorow}
\author{M. M. Fogler}
\author{L. V. Butov}
\affiliation{Department of Physics, University of California at San Diego, La Jolla, CA 92093-0319, USA}

\author{S. Hu}
\author{A.~Mishchenko}
\author{A. K. Geim}
\affiliation{School of Physics and Astronomy, University of Manchester, Manchester M13 9PL, UK}

\date{\today}% It is always \today, today,

\begin{abstract}

We report an experimental study of excitons in a double quantum well van der Waals heterostructure made of atomically thin layers of \Mo* and hexagonal boron nitride (hBN). The emission of neutral and charged excitons is controlled by gate voltage, temperature, and both the helicity and the power of optical excitation.

\end{abstract}

\maketitle

\begin{figure}[tbh]
\includegraphics[width=3.5 in, bb = 0 0 1800 1800]{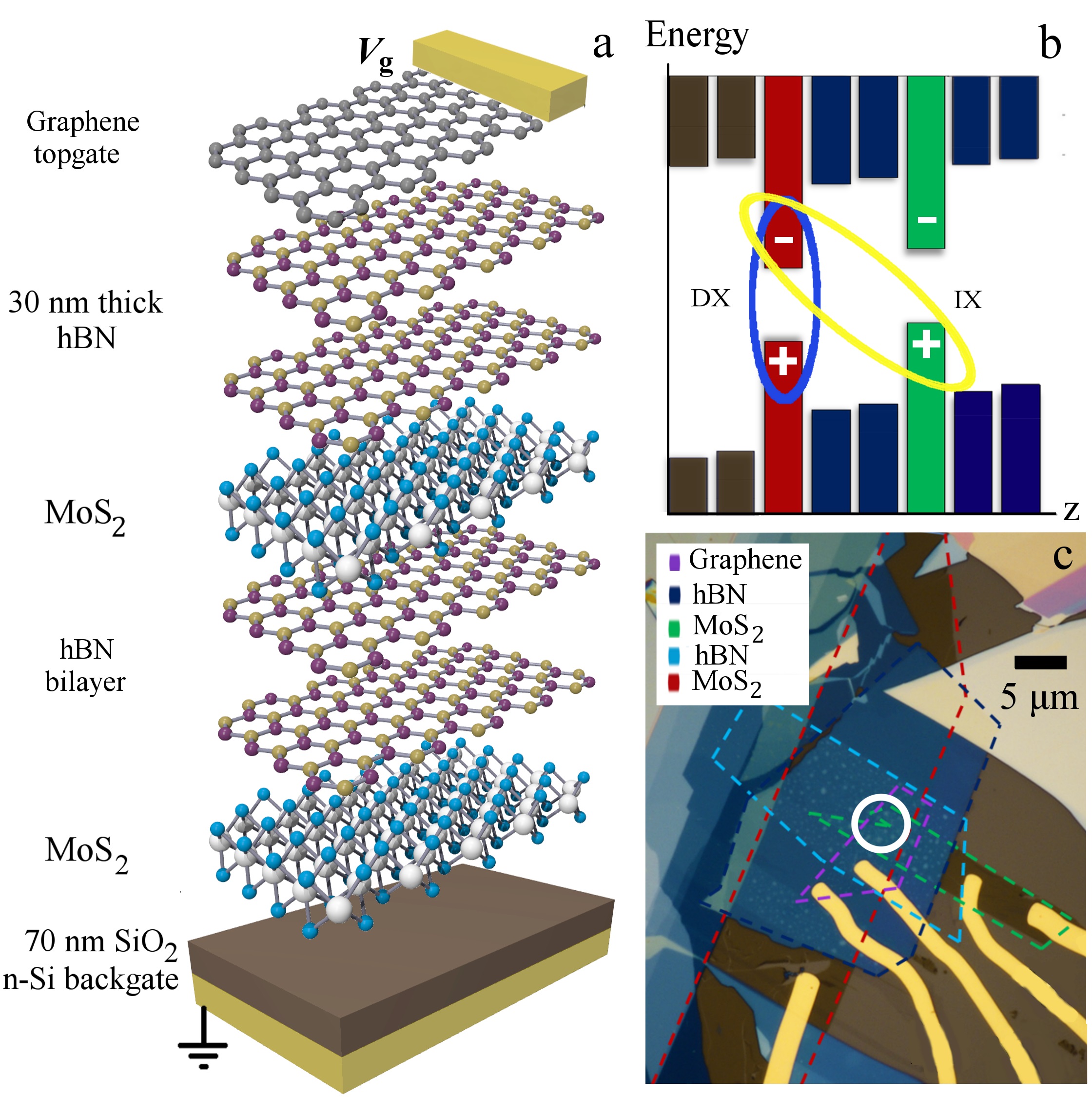}
\caption{The coupled quantum well van der Waals heterostructure. Layer (a) and energy-band (b) diagrams. The ovals indicate a direct exciton (DX) and an indirect exciton (IX) composed of an electron ($-$) and a hole ($+$). (c) Microscope image showing the layer pattern of the device. The position of the laser excitation spot is indicated by the circle.
} \label{fig:device}
\end{figure}

Van der Waals heterostructures composed of ultrathin layers of transition metal dichalcogenides (TMD), such as \Mo*, WSe$_2$, etc., offer an opportunity to realize artificial materials with designable properties, forming a new platform for studying basic phenomena and developing optoelectronic devices \cite{Geim2013vdw}. In TMD structures, excitons have high binding energies and are prominent in the optical response. The energy, intensity, and polarization of exciton emission gives information about electronic, spin, and valley properties of TMD materials~\cite{Splendiani2010epi, Mak2010atm, Zeng2012vpi, Mak2012cvp, Cao2012vsc, Cheiwchanchamnangij2012qbs, Ramasubramaniam2012lee, Wu2013etv, Mak2013tbt, Ross2013ecn, Berkelbach2013tnc, Shi2013qbs, Qiu2013osm, Ye2014ped, Chernikov2014nhe, Xu2014spl, Fang2014sic, Cheng2014epg, Zhang2014ale, Rivera2015oll, Qiu2015eos, Palummo2015erl}.

Exciton phenomena are expected to become even richer in structures that contain two 2D layers. The energy-band diagram of such a coupled quantum well (CQW) structure is shown schematically in Figure~1b. Previous studies of GaAs \cite{Islam1987ega}, AlAs \cite{Zrenner1992iec}, and InGaAs \cite{Butov1995dim} CQWs showed that excitons in these structures can be effectively controlled by voltage and light. Two types of excitons are possible in a CQW structure. Spatially direct excitons (DXs) are composed of electrons and holes in the same layer, while indirect excitons (IXs) are bound states of electrons and holes in the different layers separated by a distance $d$, Figure~1b. IXs can form quantum degenerate Bose gases~\cite{Lozovik1976nms, Fukuzawa1990pcl}. The realization and control of quantum IX gases was demonstrated~\cite{High2008cef, High2012sci} in GaAs CQW structures at temperatures $T$ below a few degrees Kelvin. In a recent theoretical work~\cite{Fogler2014hts} it was predicted that the large exciton binding energies in TMD CQW structures may bring the domain of these phenomena to high temperatures. On the other hand, DXs in TMD CQW structures have a high oscillator strength making these structures good emitters~\cite{Splendiani2010epi, Mak2010atm, Zeng2012vpi, Mak2012cvp, Cao2012vsc, Cheiwchanchamnangij2012qbs, Ramasubramaniam2012lee, Wu2013etv, Mak2013tbt, Ross2013ecn, Berkelbach2013tnc, Shi2013qbs, Qiu2013osm, Ye2014ped, Chernikov2014nhe, Xu2014spl, Fang2014sic, Cheng2014epg, Zhang2014ale, Rivera2015oll, Qiu2015eos, Palummo2015erl}. CQW structures allow control of the exciton emission by voltage. These properties make CQW structures an interesting new system for studying exciton phenomena in TMD materials.

The DX binding energy $E_\mathrm{DX}$ is larger~\cite{	Fogler2014hts} than that $E_\mathrm{IX}$ of the IXs, so in the absence of an external field the DXs are lower in energy. The electric field $F$ normal to the layers induces the energy shift $e F d$ of IXs. The transition between the direct regime where DXs are lower in energy to the indirect regime where IXs are lower in energy occurs when $e F d > E_\mathrm{DX} - E_\mathrm{IX}$~\cite{Butov1995dim}. Both direct and indirect regimes show interesting exciton phenomena. The indirect regime was considered in earlier studies of GaAs \cite{Islam1987ega}, AlAs \cite{Zrenner1992iec}, InGaAs \cite{Butov1995dim}, and TMD \cite{Fang2014sic, Rivera2015oll} CQW structures. The direct regime in TMD CQW structures is considered in this work. Exploring the direct regime is essential for understanding both the universal properties of complex exciton systems in CQW structures and the specific properties of direct excitons in TMD layers. We found that the exciton spectra in the direct regime have three exciton emission lines. The ability to control the CQW structure by voltage provides an important tool for understanding the complex exciton emission in TMD structures. The measured dependence of exciton spectra on voltage, temperature and excitation indicated that the lines correspond to the emission to neutral and charged excitons.

\begin{figure}[bth]
\includegraphics[width=3.5in,  bb = 0 0 1500 1800]{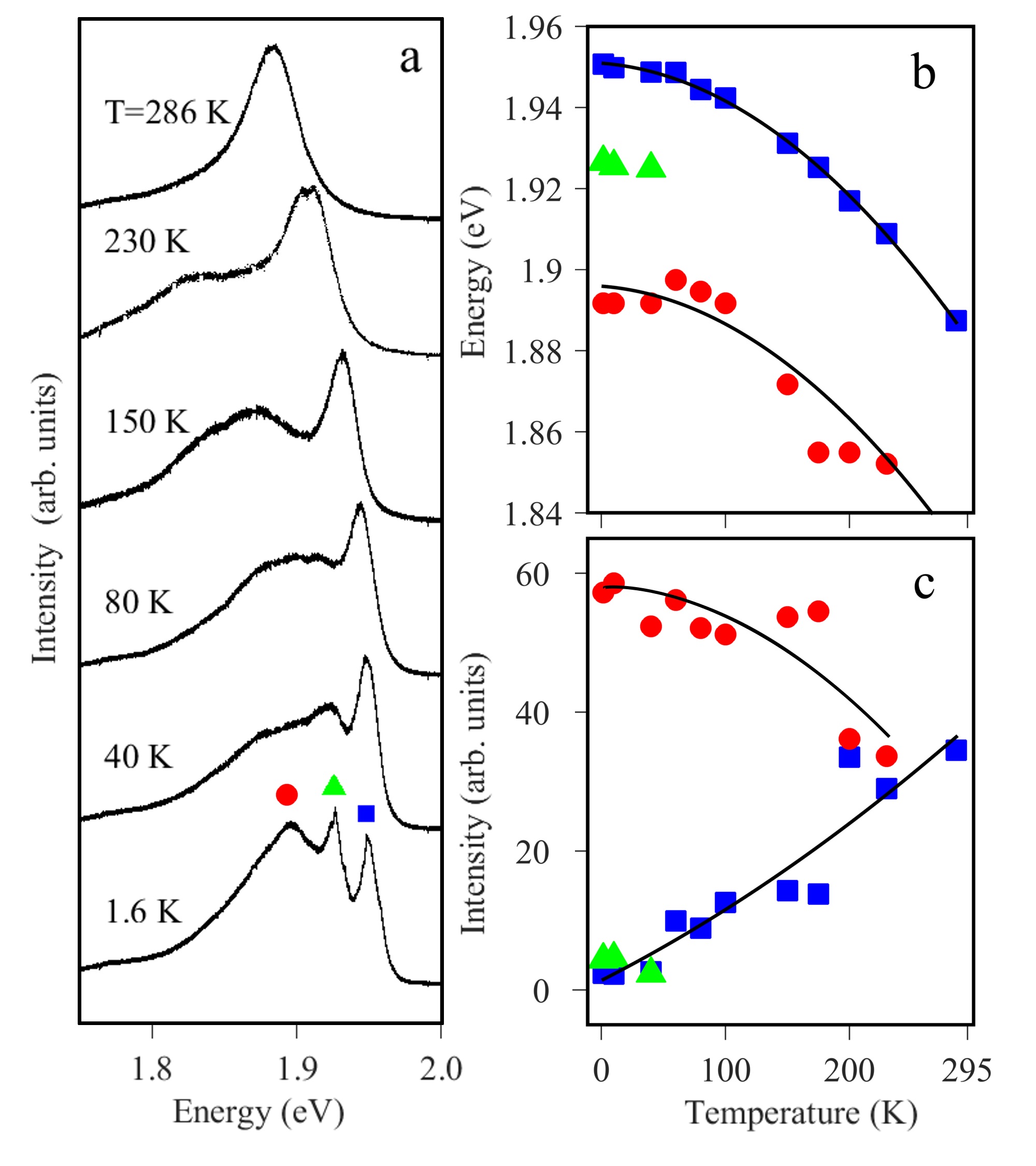}
\caption{Temperature dependence. (a) Emission spectra at different $T$. The energy (b) and intensity (c) of the emission lines marked in (a) vs. $T$. The curves are guides to the eye. $P_\mathrm{ex} = 0.8\unit{mW}$, $E_\mathrm{ex} = 3.1\unit{eV}$, $V_g = 0$.
} \label{fig:T}
\end{figure}

The structure studied here was assembled by stacking mechanically exfoliated layers on a Si/SiO$_2$ substrate, which acts as a global backgate (Figure~1a). The top view of the device showing the contours of different layers is presented in Figure~1c. The CQW is formed where the two \Mo* monolayers, separated by an hBN bilayer, overlap. The upper $20$--$30\unit{nm}$ thick hBN served as a dielectric cladding layer for a top graphene electrode. Voltage $V_g$ applied between the top graphene layer and a backgate was used to create the bias across the CQW structure.

\begin{figure}[b]
\includegraphics[width=3.5in,  bb = 0 0 1500 1800]{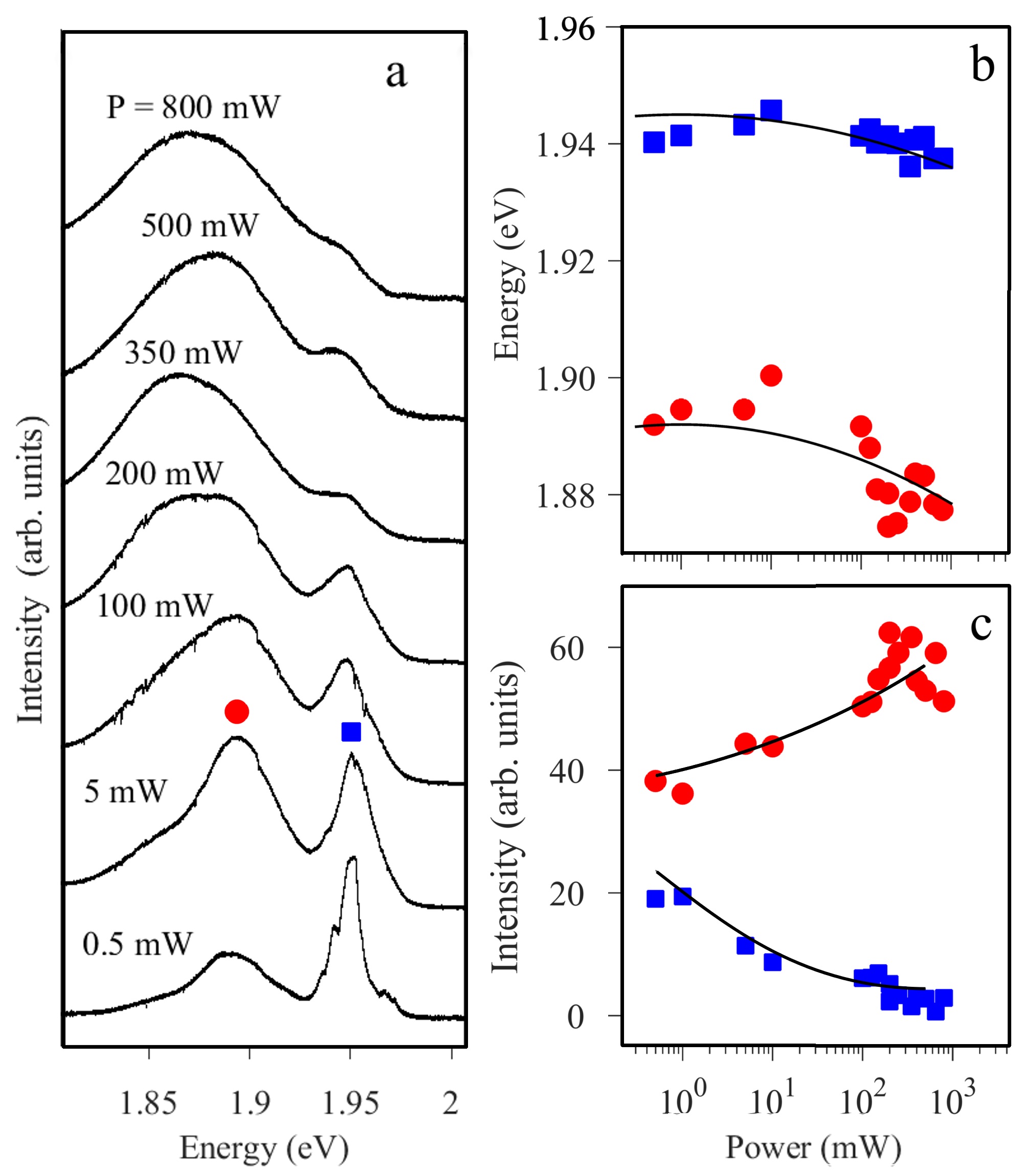}
\caption{Excitation power dependence. (a) Emission spectra at different $P_\mathrm{ex}$. The energy (b) and intensity (c) of the emission lines marked in (a) vs. $P_\mathrm{ex}$. The curves are guides to the eye. $T = 2\unit{K}$, $V_g = 0$, $E_\mathrm{ex} = 2.3\unit{eV}$.
} \label{fig:Power}
\end{figure}

The excitons were generated by continuous wave (cw) semiconductor lasers with excitation energies $E_\mathrm{ex} = 3.1$, $2.3$, or $1.96\unit{eV}$ focused to a spot of diameter $\sim 5\,\mu\mathrm{m}$ (the circle in Figure~1b). The photoluminescence (PL) spectra were measured using a spectrometer with resolution $0.2\unit{meV}$ and a CCD. In time-resolved PL measurements excitons were generated by a pulsed semiconductor laser with $E_\mathrm{ex} = 3.1\unit{eV}$ and the emitted light was diffracted by the spectrometer and detected by a photomultiplier tube and time correlated photon counting system. The measurements were performed in a $^4$He cryostat.

Figure~2 shows the PL spectra at different temperatures $T$. At the lowest $T$, the spectrum consists of two high-energy emission lines with the linewidth $\sim 20\unit{meV}$ and a broader low-energy line. Additional data and analysis presented below suggest that the high-energy lines correspond to the emission of neutral DXs while the low-energy line to the emission of charged DXs also known as trions.

The energy splitting of $25\unit{meV}$ between the high-energy emission lines constitutes only $5\%$ of the \Mo* exciton binding energy \cite{Cheiwchanchamnangij2012qbs, Ramasubramaniam2012lee, Berkelbach2013tnc, Shi2013qbs, Qiu2013osm, Ye2014ped, Chernikov2014nhe, Zhang2014ale, Qiu2015eos} of about $0.5\unit{eV}$. It is also much smaller than $0.2\unit{eV}$ energy difference of the $\mathrm{A}$ and $\mathrm{B}$ excitons~\cite{Mak2010atm} caused by the spin-orbit splitting of the valence band (see Figure~4c). These data indicate that the high-energy lines represent different species of $\mathrm{A}$ excitons. They can be $\mathrm{A}$ excitons with different electron spin states. The calculated $10\%$ difference~\cite{Kormanyos2014soc} in the masses, $0.44$ vs. $0.49m_0$, of the conduction band spin states results in a $5\%$ difference in the reduced electron-hole masses and, in turn, exciton binding energies. This leads to the energy splitting $\sim 25\unit{meV}$ consistent with the experiment. 

It is worth noting that the two \Mo* layers in the structure have inequivalent dielectric environment (Figure~1). This may lead to the difference in the binding energy of excitons in these layers in the effective mass approximation~\cite{Supporting}. However, experimental and theoretical studies show that the TMD excitonic states with large binding energy are robust to environmental perturbations~\cite{Ye2014ped}, meaning the exciton energy is the same for the two \Mo* layers in the structure.

The lower-energy emission line is shifted by about $50\unit{meV}$ from the first two (Figure~2). This shift is in the range, $20$--$50\unit{meV}$, of trion binding energies reported~\cite{Mak2012cvp, Mak2013tbt, Ross2013ecn, Berkelbach2013tnc} for monolayer \Mo*. The relative intensity of the high-energy exciton lines increases with $T$ (Figure~2), which is consistent with thermal dissociation of trions. The observed red shift of the lines with increasing temperature originates from the band gap reduction, which is typical in semiconductors,\cite{Varshni1967tde} the TMDs included~\cite{Korn2011ltp, Ross2013ecn, Soklaski2014teo, Zhang2014ale}.

Figure~3 shows the dependence of the exciton PL on the excitation power $P_\mathrm{ex}$. The relative intensity of the trion line increases with $P_\mathrm{ex}$ (Figure~3). This effect may be due to an enhanced probability of trion formation at larger carrier density. A similar increase of the trion PL intensity relative to the exciton was observed in earlier studies of GaAs CQW structures~\cite{Butov2001ces}.

\begin{figure}[t]
\includegraphics[width=3.5in,  bb = 0 0 1500 1800]{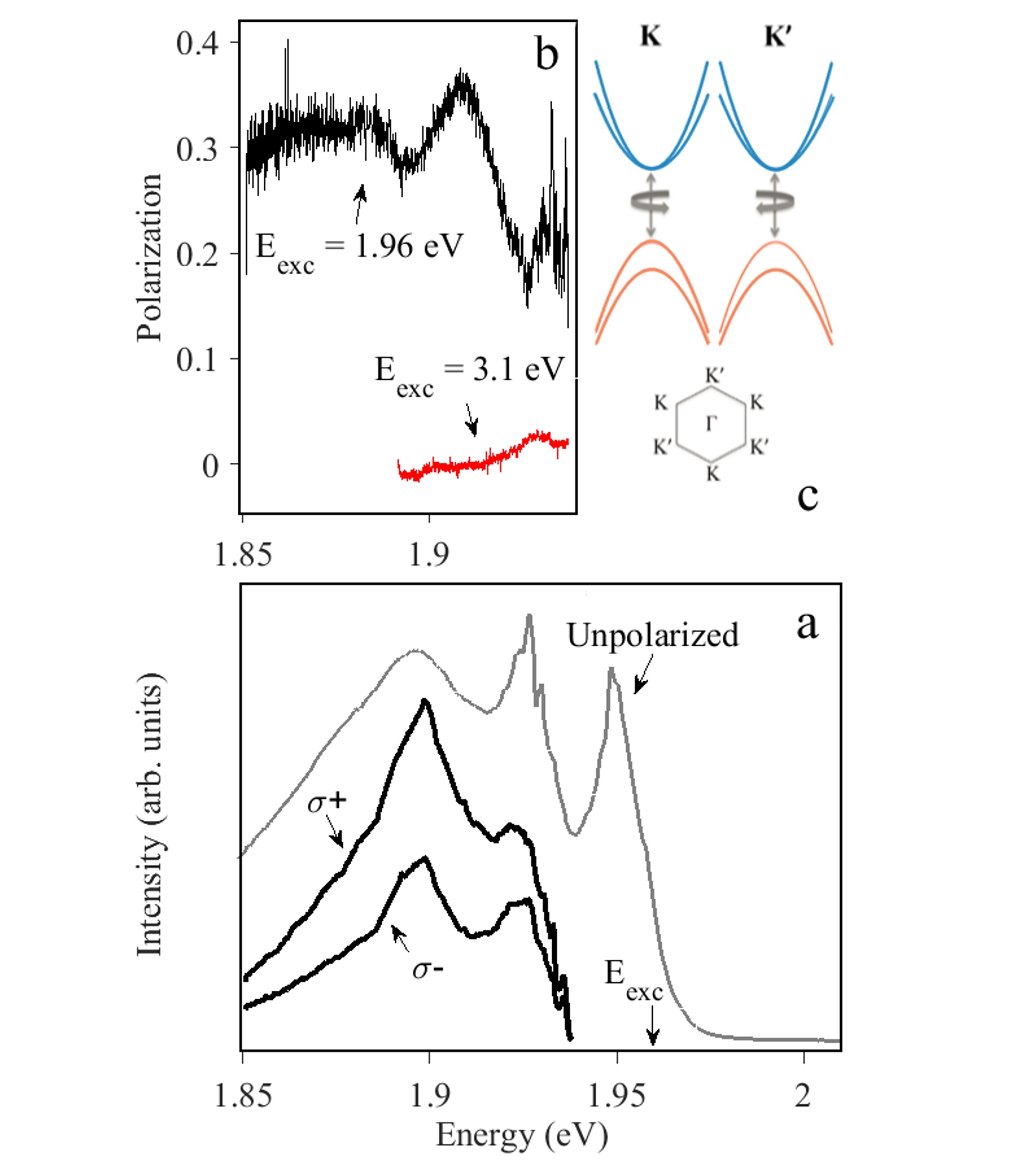}
\caption{Emission polarization. (a) Emission spectra in $\sigma^+$ and $\sigma^-$ polarizations. The laser excitation is $\sigma^+$ polarized, $P_\mathrm{ex} = 0.8\unit{mW}$, $T = 2\unit{K}$, $V = 0$, $E_\mathrm{ex} = 1.96\unit{eV}$. An unpolarized spectrum at $P_\mathrm{ex} = 1\unit{mW}$, $T = 2\unit{K}$, $V = 0$, $E_\mathrm{ex} = 3.1\unit{eV}$ is shown for comparison. (b) The emission polarization for low-energy excitation [indicated by an arrow in (a)] $E_\mathrm{ex} = 1.96\unit{eV}$ and high-energy excitation $E_\mathrm{ex} = 3.1\unit{eV}$. (c) Schematic illustrating the bands, coupling of valley and spin degrees of freedom, and optical transitions.
} \label{fig:Polarization}
\end{figure}

Figure~4 shows that the polarization of exciton emission can be controlled by the helicity of optical excitation. For a circularly polarized excitation nearly resonant with the exciton line, a high degree of circular polarization $\sim 30\%$ of exciton PL is observed (Figure~4a,b) which  is consistent withstudies of monolayer TMD~\cite{Zeng2012vpi, Mak2012cvp, Cao2012vsc, Wu2013etv, Xu2014spl}. This observation indicates that the spin relaxation time is long compared to the exciton recombination and energy relaxation times~\cite{Maialle1993esd}. The conventional explanation for the slow spin relaxation of excitons invokes spin-orbit coupling (SOC) and spin-valley coupling effects. As illustrated in Figure~4c, the SOC splits valence band of the \Mo* monolayers, leading to the appearance of the aforementioned $\mathrm{A}$ and $\mathrm{B}$ exciton states. The $\mathrm{B}$ excitons are $\sim 0.2\unit{eV}$ higher in energy and their contribution to the PL is negligible. The $\mathrm{A}$ excitons can come from either $\mathrm{K}$ or $\mathrm{K}'$ valley. It is important however that the spin and valley indices are coupled, so that exciton spin relaxation requires inter-valley scattering (Figure~4c). If this scattering is weak, the spin relaxation can be long. Virtually no circular polarization is observed for nonresonant optical excitation (Figure~4b), indicating that the high-energy photoexcited carriers loose their spin polarization during energy relaxation. Our time-resolved PL measurements revealed that the exciton and trion lifetimes are short, shorter than the $0.25\unit{ns}$ resolution of the photon counting system. Such small lifetimes facilitate the realization of the regime where the spin relaxation time is long compared to the exciton recombination time, and therefore, the polarization of exciton emission remains high.

\begin{figure}[b]
\includegraphics[width=3.5in,  bb = 0 0 1500 1800]{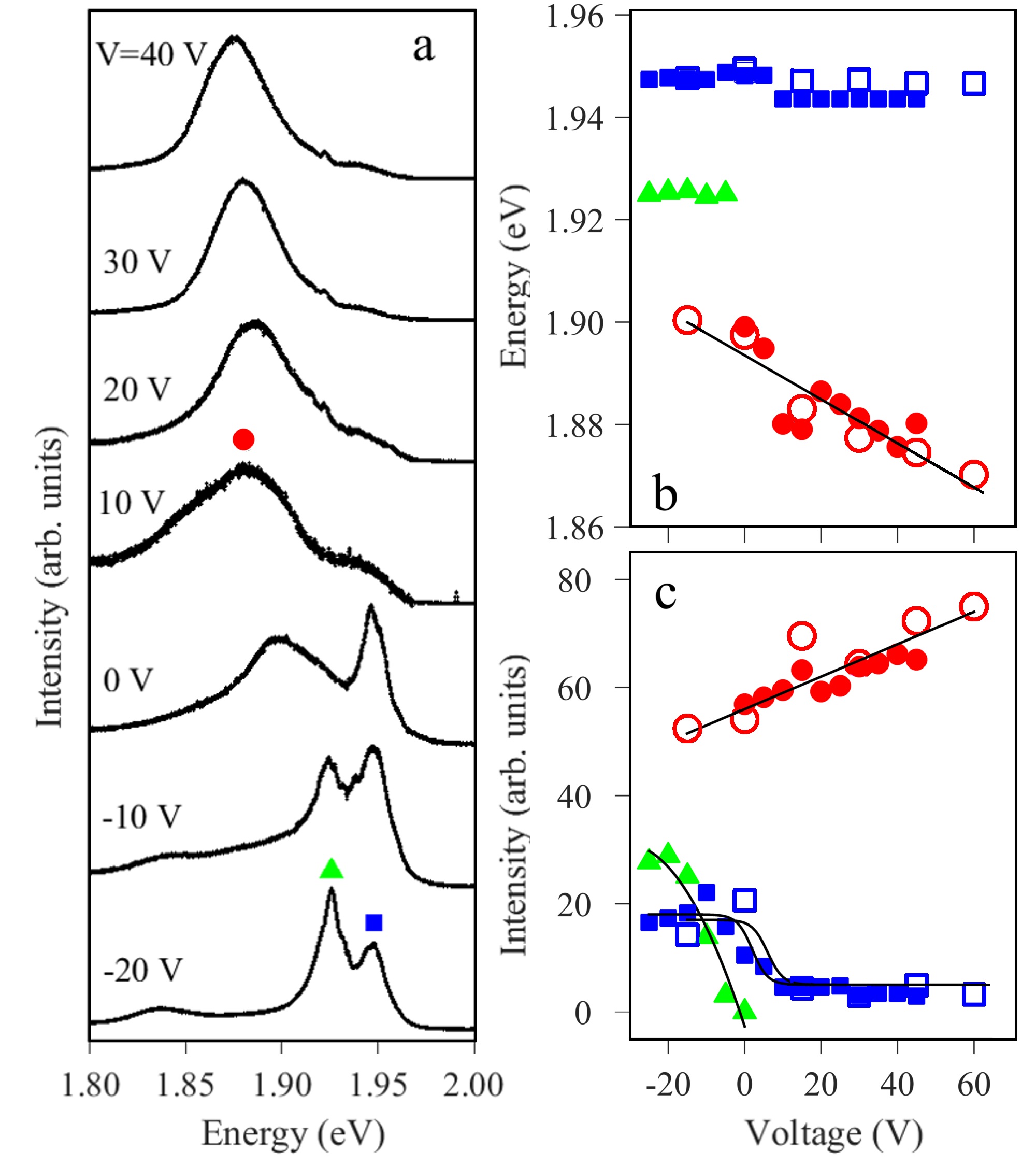}
\caption{Gate voltage dependence. (a) Emission spectra at different $V_g$. The energy (b) and intensity (c) of the emission lines marked in (a) vs. $V_g$. The curves are guides to the eye. The solid (open) symbols correspond to $E_\mathrm{ex} = 3.1 (2.3)\unit{eV}$, $P_\mathrm{ex} = 0.8\unit{mW}$, $T = 2\unit{K}$.
}
\label{fig:V}
\end{figure}

Figure~5 shows the gate-voltage dependence of the exciton PL. The small exciton lifetime $< 0.25\unit{ns}$ indicates the direct regime in the studied range of voltage because the IX lifetimes are expected to be in the ns range.~\cite{Fogler2014hts, Rivera2015oll, Fang2014sic, Palummo2015erl} The positions of the exciton lines remain essentially unchanged while the trion line exhibits a red shift with the slope $\lesssim 0.3\unit{meV}$ per $1\unit{V}$ of $V_g$. The smallness of the shifts of the lines corroborates the conclusion that the CQW is in the direct regime. Indeed, if we assume that the electric field in the device is uniform, the IX energy shift with voltage should be $\delta E_\mathrm{IX} / V_g = e F d / V_g \sim 10\unit{meV/V}$. The main effect of the gate voltage in the direct regime is the control of the exciton and trion PL intensities: the high-energy exciton emission increased at negative $V_g$, while the low-energy trion emission increased at positive $V_g$ (Figure~5). This behavior is attributed by the voltage-dependent electron concentration $n_e$ in the \Mo* layers. The initial electron concentration $n_0$ at $V_g = 0$ arises from unintentional dopants typically present in \Mo* materials. The change $\Delta n_e = n_e(V_g) - n_0$ of $n_e$ as a function of $V_g$ can be estimated from simple electrostatics. Treating the CQW as a single unit and neglecting a minor contribution from quantum capacitance, we find
\begin{equation}
\Delta n_e = \frac{C_a R_a - C_b R_b}{R_a + R_b}\,
 \frac{V_g}{e}\,,
\label{eqn:delta_n_e}
\end{equation}
where $C_{a, b}$, $R_{a, b}$ are the geometric capacitances and leakage resistances of the dielectrics above (below) this double layer. (Incidentally, the leakage current across the device did not exceed a few $\mu\mathrm{A}$ until an eventual breakdown of the device at $V_g \sim 70\unit{V}$.) Since generally $C_a R_a \ne C_b R_b$, the applied voltage changes $n_e$ and, as a result, modifies the concentration of trions relative to neutral excitons.

In summary, we presented optical studies of excitons in a \Mo* coupled quantum well van der Waals heterostructure. We observed three emission lines. The dependence of these lines on experimental parameters indicates that the two high energy lines correspond to the emission of neutral excitons and the lowest energy line to the emission of charged excitons (trions). We demonstrated control of the exciton emission by gate voltage, temperature, and also by the helicity and power of optical excitation.

This work was supported by the U.S. Department of Energy, Office of Basic Energy Sciences under award DE-FG02-07ER46449.
M.M.F. was supported by the Office of the Naval Research.
Work at the University of Manchester was supported by the European Research Council and the Royal Society.

\bibliography{MoS2_APL}
\end{document}